\documentclass[aps, prl, 10pt, superscriptaddress, twocolumn]{revtex4-1}
	\usepackage[bookmarks=false,linkcolor=blue,urlcolor=blue,colorlinks,citecolor=blue]{hyperref}

	\usepackage{graphicx}
	\usepackage{color}
	\usepackage{amsmath}
	\usepackage[normalem]{ulem}
	\usepackage[toc,page]{appendix}

\newcommand{\be}{\begin{equation}}
\newcommand{\ee}{\end{equation}}
\newcommand{\p}{\partial}
\renewcommand{\d}{\partial}
\newcommand{\la}{\label}
\newcommand{\bea}{\begin{eqnarray}}
\newcommand{\eea}{\end{eqnarray}}

\begin{document}

\title{\begin{flushright}\vspace{-1in}
			\mbox{\normalsize  EFI-17-13}
		\end{flushright}
	Transport signatures of Hall viscosity \vskip 20pt
	 }

\author{Luca V.~Delacr\'etaz}
\affiliation{Department of Physics, Stanford University,
Stanford, CA 94305-4060, USA}
\author{Andrey Gromov}
\affiliation{Kadanoff Center for Theoretical Physics 
and Enrico Fermi Institute, University of Chicago, Chicago, Illinois 60637}

\date{\today}

\begin{abstract}
Hall viscosity is a non-dissipative response function describing momentum transport in two-dimensional systems with broken parity. It is quantized in the quantum Hall regime, and contains information about the topological order of the quantum Hall state. Hall viscosity can distinguish different quantum Hall states with identical Hall conductances, but different topological order. To date, an experimentally accessible signature of Hall viscosity is lacking. We exploit the fact that Hall viscosity contributes to charge transport at finite wavelengths, and can therefore be extracted from non-local resistance measurements in inhomogeneous charge flows. We explain how to determine the Hall viscosity from such a transport experiment. In particular, we show that the profile of the electrochemical potential close to contacts where current is injected is sensitive to the value of the Hall viscosity.
\end{abstract}


\maketitle


\paragraph{Introduction.} 
Topological revolution in condensed matter physics started with the experimental observation of the precise quantization of Hall conductance in two dimensional electron systems (2DES) in strong magnetic fields. 2DES subject to a strong magnetic field enter a quantum Hall (QH) regime that is characterized by the vanishing of all dissipative response functions, and precise quantization of the Hall conductance, equal to the filling fraction $\nu$. From the theoretical standpoint Hall conductance does not uniquely characterize a QH state. Indeed, qualitatively different (fractional) QH states can appear at the same filling fraction. A notorious example is the $\nu=5/2$ plateau, whose topological order is unknown although numerous candidate states have been proposed \cite{willett1987observation, moore1991nonabelions, levin2007particle, PhysRevLett.99.236807, pakrouski2015phase, zucker2016particle}. 

One route to identifying the topological order is to look for further transport signatures that go beyond the Hall conductance. The two most prominent ones are the Hall viscosity \cite{1995-AvronSeilerZograf} and thermal Hall conductance \cite{Kane-Fisher-LR, 2000-ReadGreen, CappelliLR, Banerjee:2017qy}. The former has been studied extensively \cite{2007-TokatlyVignale, 2011-ReadRezayi, bradlyn-read-2012kubo, 2012-HoyosSon, Abanov-2014, Gromov-galilean, wiegmann-abanov-2013, klevtsov2015precise, hoyos2014hall, biswas2013semiclassical, bradlyn2014low, abanov2013FQHE} and is the subject of the present Letter. Hall viscosity $\eta_{\rm H}$ is a non-dissipative response function that describes momentum transport. It is defined as a non-dissipative component of the viscosity tensor \cite{1995-AvronSeilerZograf}
\be\la{eq_hallvisc}
T^{\rm Hall}_{ij} = \eta_{\rm H} (\epsilon_{ik} v_{kj} + \epsilon_{jk} v_{ki})\,,
\ee
where $v_{ij} = (\partial_i v_j+\partial_j v_i)/2$ is the symmetrized derivative of the velocity of the electron fluid. In a rotationally invariant QH state the Hall viscosity is proportional to a topological quantum number known as the shift $\mathcal S$ \cite{2009-Read-HallViscosity}
\be
\eta_{\rm H} = \hbar \frac{\mathcal S}{4} n\,,
\ee
where $n$ is the electron density \footnote{In the absence of rotational invariance the Hall viscosity is no longer quantized \cite{AGBB}}. The shift has been measured in a photonic integer QH system \cite{schine2015synthetic}. The dimensionless kinematic Hall viscosity $\eta_{\rm H}/\hbar n$ is, therefore, quantized. Unlike the Hall conductance, the Hall viscosity (or the shift) is a bulk property of the topological phase, and, in general, is {\it not} encoded in the properties of gapless edge modes \cite{gromov2016boundary}. Thus, in order to measure the Hall viscosity a bulk transport experiment is required. This is not in contradiction with the insulating nature of the QH regime -- in a typical transport experiment both bulk and metallic edges carry electric current \cite{thouless1993edge, wexler1994current}.
 It is worth emphasizing that (unlike shear and bulk viscosities) the sign of the Hall viscosity is not restricted by the second law of thermodynamics. Indeed, for the Pfaffian and anti-Pfaffian states -- promising candidates for the observed $5/2$ state -- the Hall viscosity has opposite signs \cite{2011-ReadRezayi}.

Hall viscosity contributes to electromagnetic transport through a finite wavelength correction to the Hall conductivity \cite{2012-HoyosSon, bradlyn-read-2012kubo}
\be
\sigma_{\rm H}(k) = \sigma^{(0)}_{\rm H} + \sigma^{(2)}_{\rm H} |k\ell|^2 + \ldots\,,
\ee
where $\sigma^{(0)}_{\rm H} = \nu/2\pi$ \footnote{From now on we will work in the natural units $e=\hbar=c=1$.}, $\ell$ is the magnetic length and for non-dissipative Galilean invariant systems \cite{2012-HoyosSon}
\be\la{sigma2}
\sigma^{(2)}_{\rm H} = \frac{\nu}{2\pi} \frac{\eta_{\rm H}}{n} + m^\star \kappa^{-1}_{\rm int}\,.
\ee
Here $m^\star$ is the effective electron mass and $\kappa^{-1}_{\rm int}$ is the inverse internal compressibility \cite{bradlyn-read-2012kubo}, related to the magnetic susceptibility by $\kappa^{-1}_{\rm int} = B^2 \chi$. This second term is a thermodynamic property $ \kappa^{-1}_{\rm int} = B^2(\partial^2 \varepsilon/\partial B^2)_\nu$, where $\varepsilon$ is the energy density.

 The first term in Eq. \eqref{sigma2} is universal, while the second term is not. However, in the limit of vanishing LL mixing (which can be achieved taking $m^\star \rightarrow 0)$ $m^\star \kappa^{-1}_{\rm int}$ tends to a universal value $\nu/2\pi$. Thus we find 
\be\la{eq:est}
\sigma^{(2)}_{\rm H} = \frac{\nu}{2\pi} \left( \frac{\eta_{\rm H}}{n} + 1 \right)\,.
\ee

Eq.~\eqref{sigma2} holds for non-dissipative Galilean invariant systems in a magnetic field. It can be easily generalized to dissipative systems with or without Galilean symmetry -- this is done below in the hydrodynamic section. 
Thus, although we are mostly interested in the QH regime, our derivations will hold both in dissipative and non-dissipative regimes.

{ We emphasize that in order to take full advantage of our proposal, the susceptibility $\kappa^{-1}_{\rm int}$ must be either calculated or measured independently. In the IQH case the estimate of Eq.~\eqref{eq:est} should be reliable, but with the ultimate objective of studying fractional states in mind, an experimental protocol is desired. The magnetic susceptibility can be determined from the local distribution of magnetization currents in the presence of an inhomogeneous magnetic field \cite{Halperin-heat}, which would require imaging of the local current profile. An alternative experimental route is to measure the optical conductivity at finite wavelengths. Indeed, although $\kappa_{\rm int}^{-1}$ and $\eta_{\rm H}$ enter in the fixed combination \eqref{sigma2} in dc electric transport, the hydrodynamic treatment below shows that this is no longer the case at finite frequency (see also Ref.~\cite{bradlyn-read-2012kubo}). Finally, any experimental measurement of the equation of state of the 2DES will lead to the determination of $\kappa_{\rm int}^{-1}$. 
} 

\begin{figure}[h]
\centerline{
\includegraphics[width=\linewidth,angle=0]{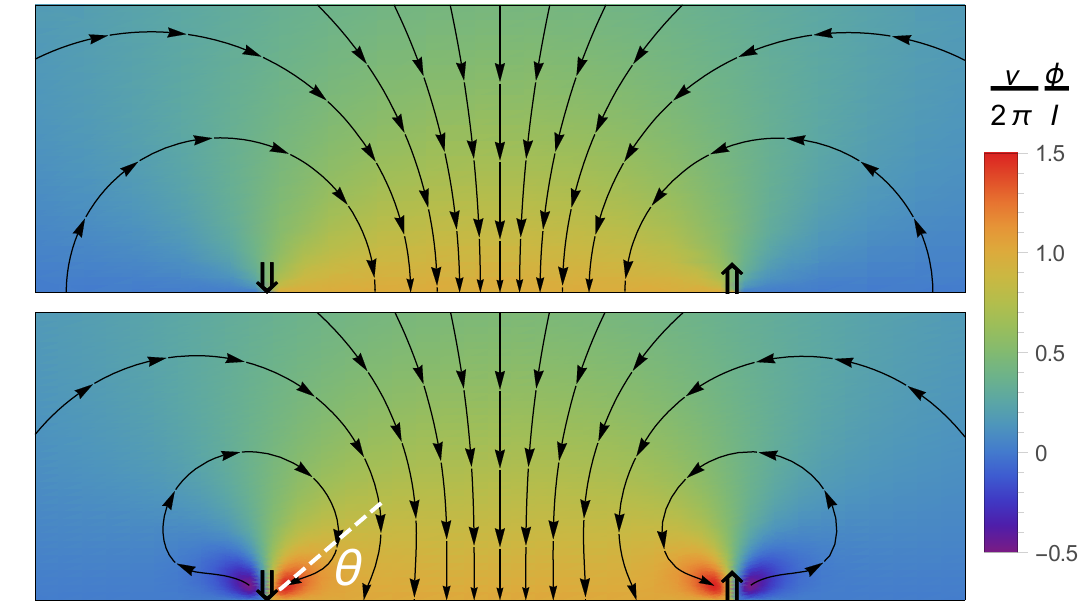}}
\caption{\label{fig_flow} {\bf Top}, current is injected through the right contact ($\Uparrow$) and drained in the left contact ($\Downarrow$). The color indicates the magnitude of the electrochemical potential and the arrows indicate its gradient, in the absence of $\rho_{\rm H}^{(2)}$. {\bf Bottom}, same setup in the presence of $\rho_{\rm H}^{(2)}$, which affects the potential near the contacts. $\rho_{\rm H}^{(2)}$ can be extracted from a measurement of the potential along the white dashed line $y=x\tan \theta$, shown in Fig.~\ref{fig_finite_contact}.}
\end{figure}

The qualitative effect of the Hall viscosity is demonstrated in Fig.~\ref{fig_flow}. In this setup there is a small current source and a drain located at the lower edge of the system. The electrochemical potential acquires an extra contribution that leads to a pronounced maximum of the potential on one side of the contact and local minimum on the other side. These corrections decay as $1/r^2$, where $r$ is the distance from the contact. The coefficient in front of this extra correction is sensitive to $\sigma^{(2)}_{\rm H}$, from which the Hall viscosity can be determined up to the details discussed in the Introduction. The spatial distribution of the electric current density is identical for both cases (in agreement with \cite{ganeshan2017odd}) in Fig.~\ref{fig_flow}, however the resulting distribution of electrochemical potential is affected by the Hall viscosity.
%

\paragraph{Ohm's Law.} 
 
Within linear response, a small applied DC electric field will source a DC electric current
\begin{equation}\label{eq_ohm}
E_i(k) = \rho_{ij}(k)j_j(k)\,,
\end{equation}
where $\rho_{ij} \equiv (\sigma^{-1})_{ij}$.
At distances long compared to the magnetic length $\ell$, Eq.~\eqref{eq_ohm} can be written in a gradient expansion 
\bea\nonumber
E_i(x) &=& \left[\rho^{(0)} - \rho^{(2)}(\ell \nabla)^2\right] j_i(x) \\ \label{eq_pde1}
	& +& \left[\rho_{\rm H}^{(0)} - \rho_{\rm H}^{(2)}(\ell \nabla)^2\right]\epsilon_{ij} j_j(x)+ \ldots\, ,
\eea
where we have used the continuity equation $\p_i j_i =0$ to eliminate some of the terms. This differential equation can be interpreted as a (linearized) Navier-Stokes equation and is derived in a hydrodynamic framework below.

Quantum Hall systems are non-dissipative at low temperatures, and the coefficients $\rho^{(2n)}$ vanish as $\exp{\left(-\frac{\Delta}{T}\right)}$, where $\Delta$ is the spectral gap -- these terms are however finite for gapless systems. They will be kept finite for some of the calculations, but will be taken to zero when discussing the QH regime, keeping the ratio $\rho^{(2)}/\rho^{(0)}$ fixed. In this limit the Hall resistance and its gradient correction are given by
\be
\rho^{(0)}_{\rm H} = -1/\sigma^{(0)}_{\rm H}\,,\qquad \rho^{(2)}_{\rm H} =  \sigma^{(2)}_{\rm H}/\left(\sigma^{(0)}_{\rm H}\right)^2\,.
\ee

The continuity equation implies that electric current can be written as a curl of a ``stream function'' $j_i=- \epsilon_{ij}\nabla_j \psi$. Writing $E=-\nabla \phi$, the divergence and curl of Eq.~\eqref{eq_pde1} give 
\begin{subequations}\label{eq_pde2}
\begin{align}
\Delta \left[\Delta - \frac{1}{\mathrm{r_H}\ell^2}\right]\psi 
	&= \frac{\Delta \phi}{\ell^2 \rho^{(2)}_H}\, , \\
\Delta \left[\Delta - \frac{1}{\mathrm{r}^2\ell^2}\right]\psi 
	&= 0\, , \label{eq_pde2b}
\end{align}
\end{subequations}
here we have introduced the dimensionless ratios $\mathrm{r}^2 =\rho^{(2)}/\rho^{(0)}$ and $\mathrm{r}_{\rm H} =  \rho^{(2)}_{\rm H}/\rho^{(0)}_{\rm H}$.  The second equation was studied in great detail by Levitov and Falkovich in Refs.~\cite{levitov2016electron, falkovich2016linking}. In the limit of weak LL mixing and for $\nu <1$ the ratio $\mathrm{r}_{\rm H}$ is given by
\be\la{eq_rH}
\mathrm{r}_{\rm H} =  \frac{\eta_{\rm H}}{n} + 1\,.
\ee
In general, $\mathrm{r}_{\rm H}$ is not universal and depends on the magnetization via \eqref{sigma2}. An expression for ${\rm r}^2$ is given in the hydrodynamics section below -- in the dissipative regime it is proportional to the ratio of the shear viscosity to the momentum relaxation rate. Although the qualitative effect depicted in Fig.~\ref{fig_flow} is generic in the presence of a non-zero $\sigma^{(2)}_{\rm H}$, the amplitude of the correction is sensitive to $\mathrm{r^2}$, which should be measured independently. In the limit when the LL mixing is neglected $\mathrm{r}$ is large.

Eqs.\,\eqref{eq_pde2}  require two boundary conditions at each boundary. The current normal to boundaries $j_n$ will be fixed, and the current along boundaries $j_l$ will be taken to satisfy ``partial-slip'' boundary conditions \cite{torre2015nonlocal}
\begin{equation}\label{eq_bc_part}
\d_n j_l = -\frac{1}{\ell_s} j_l\, ,
\end{equation}
where $n$ ($l$) is the direction normal (longitudinal) to the boundary pointing outwards. The limits $\ell_s \rightarrow 0$ and $\ell_s \rightarrow \infty$ leads to no-slip and no-stress boundary conditions respectively. We will see that in the QH regime $\ell_s$ will have to be taken to be infinite in order to agree with known features of the quantized Hall conductance.

\paragraph{Pipe flow.} 
A simple observable effect of the Hall viscosity can be seen in a pipe flow setup. The PDEs \eqref{eq_pde1} can be easily solved for a pipe flow $j_i(x,y)=(j_x(y),0)$, with $-w/2<y<w/2$. Although boundary conditions need to be imposed to determine the spatial distribution of electric current and potential, there is a specific combination of resistivities which is independent of boundary conditions:
\begin{equation}
\frac{\rho_{xx}}{\rho^{(2)}} - \frac{\rho_{xy}}{\rho_{\rm H}^{(2)}} = \frac{1}{{\rm r}^2} - \frac{1}{{\rm r_H}}\, ,
\end{equation}
as can be seen by integrating \eqref{eq_pde1} over $y$. Here the average resistivities are defined as $\langle E_i\rangle = \rho_{ij} \langle j_j\rangle$, where $\langle A \rangle\equiv \frac{1}{w}\int_{-w/2}^{w/2}A\, dy$.

Boundary conditions must be imposed in order to find the individual resistivities.
Using partial-slip boundary conditions \eqref{eq_bc_part} at both edges, the Hall resistance $R_{xy}=\rho_{xy}$ is found to be
\begin{equation}
R_{xy} = \rho_{\rm H}^{(0)} \left[1 + \frac{2\mathrm{r}_{\rm H}\ell^2}{\mathrm{r}w\ell{\rm coth\,} \frac{w}{2\mathrm{r}\ell} - 2\mathrm{r}^2\ell^2 + w \ell_s} \right]\,.
\end{equation}
It is instructive to consider a few limiting cases. First, we consider the limit of large $\mathrm{r}^2$ (which amounts to neglecting momentum relaxation) 
\be\la{moore}
R_{xy} = \rho_{\rm H}^{(0)}\left[1 + \mathrm{r_H} \frac{12\ell^2}{w(w+6\ell_s)} \right]\,.
\ee
In the no-slip limit $\ell_s \rightarrow 0$ \eqref{moore} reduces to an expression similar to the one obtained in Ref.~\cite{scaffidi2017hydrodynamic}, with the difference that the Hall viscosity is replaced by $\mathrm{r_H}$ \footnote{The contribution of $\kappa^{-1}_{\rm int}$ in Eq.\eqref{sigma2} can be neglected for Fermi liquids in a weak magnetic field, as done in \cite{scaffidi2017hydrodynamic}. Generically, however, both contributions are expected to be of the same order.}. Here we see more generally that this result is sensitive to boundary conditions through $\ell_s$, and to momentum relaxation through ${\rm r}^2$.

Next we consider the case of large $\ell_s$ (``almost no-stress'' boundary conditions). We find
\be
R_{xy} = \rho_{\rm H}^{(0)}\left[1 + \mathrm{r_H} \frac{2\ell^2}{w\ell_s}  + O(\ell_s^{-2})\right]\,.
\ee

In the QH regime, the Hall conductance in a strip geometry has an exponentially small width dependence \cite{niu1987quantum}. In order to recover this property we will fix no-stress boundary conditions when addressing the QH regime. Only in this case does one find $R_{xy} =  \rho_{\rm H}^{(0)}$, regardless of the value of $\mathrm{r}^2$, and quantization of Hall conductance is precise. A correction that is sensitive to ${\rm r_H}$ can nevertheless be observed by measuring the bulk potential in slightly more complex flows, as shown in the following section.

\paragraph{Point contacts.}
The Hall viscosity contributes to a higher gradient correction to transport -- as such its effect is most dramatic on length scales that approach the magnetic length $\ell$. This observation motivates a study of the solution to \eqref{eq_pde2} near contacts where current is injected. We consider a half-plane geometry with two contacts (see Fig.~\ref{fig_flow}), which amounts to imposing the boundary condition
\begin{equation}\label{eq_bc_inject}
j_y(x,0) = \d_x \psi(x,0) = I \left[-\delta(x) + \delta(x-w)\right]\,,
\end{equation}
where $I$ is the total current injected. The finite distance between the contacts is useful in the intermediate calculations as a long-distance regulator, but does not affect the potential and the current distributions near the contacts, so we will take $w\to \infty$ in the end. Fourier transforming the stream function along $x$, \eqref{eq_pde2} turns into an ordinary differential equation for $\psi_k(y)=\int dx\, e^{-ikx}\psi(x,y)$ which has the solution
\begin{equation}
\psi_k(y) = a_k e^{-|k|y} + b_k e^{-q y}\, ,
\end{equation}
where we imposed that the current vanish at $y\to \infty$, and $q^2 = k^2 + \ell^2\mathrm{r}^2$. The integration constants $a_k,\, b_k$ are fixed by imposing the boundary conditions \eqref{eq_bc_part} and \eqref{eq_bc_inject}. The potential can then be determined through Eq.~\eqref{eq_pde1}
\begin{equation}\label{eq_phi_cor}
\phi(x,y) = -\rho_{\rm H}^{(0)} \left[\mathrm{r}_{\rm H}\ell^2\Delta - 1\right]\psi(x,y) + \phi_0\, ,
\end{equation}
where $\phi_0$ is a constant that we drop in the following. The explicit form of the potential is given in the Supplementary Material. 

We focus in this section on signatures of the Hall viscosity in QH phases, where the appropriate boundary condition is no-stress $\ell_s=\infty$ \footnote{Different regions of parameter space (which may be relevant to systems other than QH) are studied in the Supplementary Material and summarized in Table 
\ref{tab_sum}.}. When $\ell_s<\infty$, the potential at the boundary $y=0$ receives corrections from the Hall viscosity through ${\rm r_H}$ -- however these corrections vanish when $\ell_s = \infty$. This leads, on general grounds, to the robust quantization of the Hall conductance in more complex geometries involving current contacts. For our purposes, it means that the Hall viscosity can only be extracted by measuring the bulk potential. Near the drain at $(x,y)=(0,0)$, it is given by
\begin{equation}\label{eq_R_inject}
\phi(x,y) = \rho_{\rm H}^{(0)} I \left[\frac{\pi -\theta}{\pi} +  {\rm r}_{\rm H}  \ell^2\frac{2}{\pi}\frac{xy}{(x^2+y^2)^2} + \ldots \right]\, ,
\end{equation}
where $\theta = {\rm Arg}\, (x+i y)$, and $\ldots$ denotes terms that are subleading when $\ell^2 \ll x^2 +y^2 \ll w^2 $. In deriving \eqref{eq_R_inject} we assumed that the length scale ${\rm r}\ell$ associated with the dissipative regulator is large (of order system size or $w$): the relevance of this assumption will be motivated by the hydrodynamic approach below. 

\begin{figure}[h]
\centerline{
\includegraphics[width=\linewidth,angle=0]{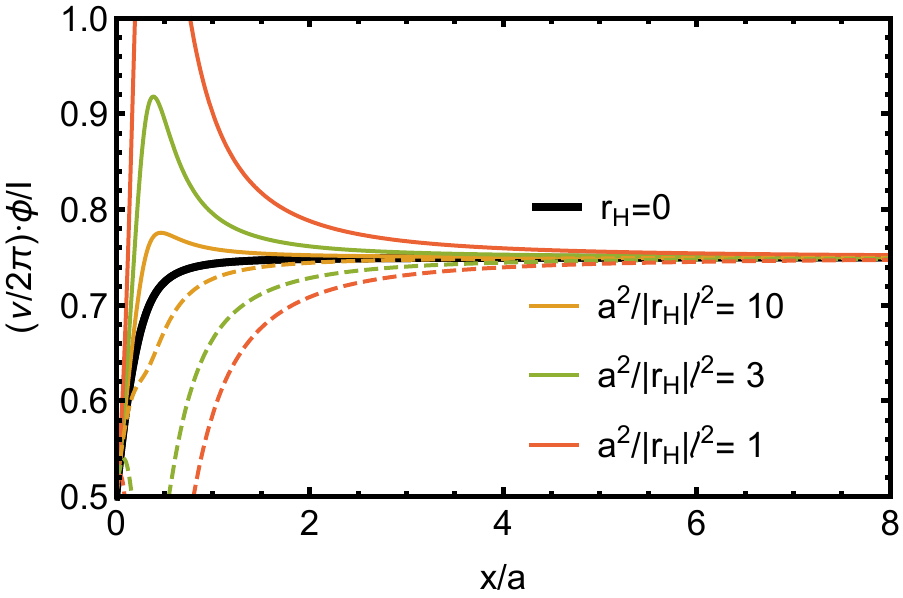}}
\caption{\label{fig_finite_contact}Electrochemical potential near the current drain, along the line $y=x\tan \theta$ shown in Fig.~\ref{fig_flow} with $\theta=\pi/4$. When $\mathrm{r_H}$ is positive (solid curves) the potential overshoots before settling to its asymptotic value $\frac{\nu I}{2\pi} \phi(x,y) \to (\pi-\theta)/\pi$ as $x\to \infty$. When  $\mathrm{r_H}$ is negative (dashed curves) the situation is reversed. The effect becomes more dramatic as the contact width $a$ is reduced to be of order $|{\rm r_H}|^{1/2}\ell$.}
\end{figure}

The correction proportional to ${\rm r_H}$ to the potential grows as one approaches the contacts. The singular behavior as $x,y\to 0$ is smeared by the finite size $a$ of the contacts (see the Supplementary Material), the resulting regulated potential profile is shown in Fig.~\ref{fig_flow}. The correction vanishes exactly at the boundary of the sample, where the potential takes its quantized value. In order to extract the Hall viscosity, the potential in the bulk must be measured, for example along the line $y=x \tan \theta$ for $0<\theta<\pi/2$. This potential is plotted in Fig.~\ref{fig_finite_contact} for $\theta=\pi/4$ (the potential profiles at other angles are similar as long as $\theta$ is not too close to $0$ or $\pi/2$). The potential profile depends qualitatively on the sign of ${\rm r_H}$.

In addition to smearing the potential near the drain, the finite size $a$ of the contacts gives a small correction in the bulk which competes with the correction due to ${\rm r_H}$. Specifically, the bulk potential still has the form \eqref{eq_R_inject}, with
\begin{equation}
{\rm r_H \ell^2} \to
{\rm r_H \ell^2} - \alpha a^2\, ,  
\end{equation}
where $\alpha$ depends on the current profile in the contacts ($\alpha=1/24$ for rectangle contacts, and $\alpha = 1/4$ for Gaussian contacts).

This procedure provides a rather robust measurement of the Hall viscosity, because (i) several measurements of the potential can be performed on a single sample, varying the distance to the contact, and (ii) the signature survives even if the contact is greater than $|{\rm r_H}|^{1/2}\ell$, as shown in Fig.~\ref{fig_finite_contact}.

\paragraph{Hydrodynamic approach.}
Eq.~\eqref{eq_pde1} can be interpreted as a linearized Navier-Stokes in the presence of weak momentum relaxation $\Gamma$
\begin{equation}\label{eq_NS}
\dot P^i + \d_j T^{j i } = F^{i\mu}J_\mu-\Gamma P^{i}\,,
\end{equation}
where $P^i$ is the momentum density of the fluid.
The linearized constitutive relation for the stress-tensor is
\begin{equation}\label{eq_T_consti}
T_{ij} = \left[p_{\rm int}  - \chi_\Omega \Omega\right]\delta_{ij} + T^{\mathrm{bulk}}_{ij} + T^{\mathrm{shear}}_{ij} + T^{\mathrm{Hall}}_{ij}+ \ldots \, ,
\end{equation}
where $\ldots$ denote higher derivative terms, $p_{\rm int} = p - MB$ is the internal pressure \cite{Halperin-heat}, $\Omega = \nabla\times v$ is the vorticity and $\chi_\Omega$ is the corresponding susceptibility \cite{khalatnikov1989introduction, jensen2012parity, kaminski2014nonrelativistic, wiegmann-abanov-2013}. The Hall contribution to the stress-tensor was given in \eqref{eq_hallvisc}. The other two contributions are dissipative and given by
\begin{equation}
T^{\rm bulk}_{ij} = \zeta \delta_{ij} (\nabla\cdot v)\, , \quad
T^{\rm shear}_{ij} = \eta \bigl[\nabla_i v_j + \nabla_j v_i - \delta_{ij} (\nabla\cdot v)\bigr]\, .
\end{equation}
We will search for stationary solutions to \eqref{eq_NS}. The continuity equation implies that the flow is incompressible $\nabla\cdot v = 0$. The Navier-Stokes equation then becomes \footnote{The assumption of stationary flow forces $\eta_H$ and $\chi_{\Omega}$ to appear in the combination $\eta_H-\chi_{\Omega}$. They are however independent parameters in general (see Eq.~\eqref{eq_T_consti}), and can be probed seperately in non-stationary flows through the optical conductivity $\sigma(\omega,k)$ \cite{bradlyn-read-2012kubo}.}
\be \la{eq_NS2}
\Gamma P_i - \eta \nabla^2 v_i + (\eta_{\rm H}-\chi_\Omega) \nabla^2 \epsilon_{ij} v_j= n (E_i + B \epsilon_{ij} v_j) -\nabla_i p_{\rm int}\,.
\ee
Momentum and velocity $P^{i}=n m^\star v^i$ are related by the effective mass density $m^\star n$ (note that in the LLL limit $m^\star \rightarrow 0$ the momentum density vanishes \cite{geracie2014hydrodynamics}). In the Galilean limit, the velocity is related to the current by $j^i = n v^i$ \footnote{Without Galilean invariance, higher derivative terms such as $\nabla^2 v_i$ can appear in the current constitutive relation. These will give additional $k^2$ corrections to charge transport, which are degenerate with the contributions from the viscosities $\eta$ and $\eta_{\rm H}$ in \eqref{eq_Ohm_hydro}.
} which can be reinserted in \eqref{eq_NS2} to recast the Navier-Stokes equation in the form of Ohm's law
\begin{equation}\label{eq_Ohm_hydro}
E_i = \left[\frac{\Gamma m^\star}{n} - \frac{\eta}{n^2}\nabla^2\right] j_i 
	- \left[ \frac{B}{n} - \frac{\eta_{\rm H} -\chi_\Omega}{n^2}\nabla^2 \right] \epsilon_{ij} j_j + \ldots\, .
\end{equation}
Comparison with \eqref{eq_pde1} gives the coefficients $\rho^{(2n)}$ and $\rho_{\rm H}^{(2n)}$. This is a generalization of \eqref{sigma2} in the presence of dissipation. Galilean invariance requires variations in the vorticity and magnetic field to appear in the combination $m^\star \delta \Omega - \delta B$ \cite{jensen2014aspects, kaminski2014nonrelativistic}. Applying this to Eq.~\eqref{eq_T_consti}, noting that $\delta p_{\rm int} = B\delta M = \frac{1}{B} \kappa^{-1}_{\rm int} \delta B$, this implies
\begin{equation}
\chi_\Omega = - \frac{m^\star}{B} \kappa^{-1}_{\rm int}\, ,
\end{equation}
so that the expression for $\rho_{\rm H}^{(2)}$ entering \eqref{eq_Ohm_hydro} exactly agrees with Ref.~\cite{2012-HoyosSon} when the dissipation is turned off \footnote{Without Galilean invariance $\chi_\Omega$ is in general independent from $\kappa^{-1}_{\rm int}$ (see also \cite{Note6}).}.

\paragraph{Conclusions.}

We have investigated how the Hall viscosity affects charge transport. Its contribution to finite wavelength corrections to the Hall conductivity lead to transport signatures in inhomogeneous setups. In particular, we have shown that the potential distribution in the QH regime acquires extra features -- demonstrated in Fig. \ref{fig_flow} and Fig. \ref{fig_finite_contact} -- close to the current injector. 
In the dissipative regime, the Hall conductance acquires a finite size correction in a pipe flow setup, which is sensitive to the boundary conditions.  
Finally, we have shown that our results can be derived from a hydrodynamics approach to QH.

\paragraph{Acknowledgments.}

It is a pleasure to thank A.~Abanov, B.~Bradlyn, D.~Feldman, M.~Geracie, K.~Jensen, S.W.~Kang, S.~Kivelson, M.~Levin, D.T.~Son and P.~Wiegmann for useful discussions. We also thank Sean Hartnoll and Andy Lucas for helpful comments on a draft of this paper. Finally, we thank T.~Scaffidi and J.~Moore for explaining the results of Ref.~\cite{scaffidi2017hydrodynamic} to us. L.V.D. was supported by the Swiss National Science Foundation, and by a DOE Early Career Award (Sean Hartnoll).  A.G. was supported by the Leo Kadanoff fellowship and the NSF grant DMS-1206648.

\paragraph{Note added.}---When the present work was partially complete we learned about Ref.\,\cite{scaffidi2017hydrodynamic} where the effect of the Hall viscosity on the flow of an electron fluid through a pipe in a weak magnetic field was investigated. Their results in the hydrodynamic regime can be recovered from \eqref{moore} by imposing no-slip conditions $(\ell_s \to 0)$ and ignoring the $m^\star \kappa^{-1}_{\rm int}$ contribution to $\rho_{\rm H}^{(2)}$ \cite{Note6}.

\bibliography{Bibliography}

\newpage

\begin{widetext}

\section*{Supplementary Material for \\ Transport signatures of Hall viscosity}

\subsection{Flow equation}
The differential equation \eqref{eq_pde2b} can be solved by Fourier transforming the stream function in the infinite $x$-direction $\psi_k(y) = \int dx\, e^{-ikx}\psi(x,y)$ \cite{levitov2016electron}, leading to
\begin{equation}\label{eq_pde_k}
(\d_y^2 - k^2)(\d_y^2 - q^2)\psi_k(y)=0\, ,
\end{equation}
with $q^2 = k^2 + \ell^2 {\rm r}^2$. After imposing the boundary conditions \eqref{eq_bc_part} and \eqref{eq_bc_inject} at $y=0$, and that the current vanish as $y\to \infty$, the unique solution is
\begin{equation}\label{eq_psi_sol}
\psi_k(y) = -I\,  \frac{1-e^{-ikw}}{ik} \left[a_k e^{-|k|y} + b_k e^{-qy}\right]\, ,
\end{equation}
with
\begin{equation}
b_k = -\frac{|k| - \ell_s k^2}{(q-|k|) - \ell_s (q^2 - k^2)}
\end{equation}
and $a_k=1-b_k$. The only purpose of keeping a finite distance $w$ between the contacts is to provide an IR regulator when \eqref{eq_psi_sol} is Fourier transformed back into real space. We will omit it in the following for readability, taking $({1-e^{-ikw}})/{ik}\to 1/ik$, and reintroduce it when performing $k$ integrals. The potential can be found from the stream function using \eqref{eq_pde1}, which in the dissipationless limit $\rho^{(2n)}\to 0$ gives
\begin{equation}\label{eq_phi_k}
\phi_k(y) = \frac{\rho_{\rm H}^{(0)}I}{ik} \left[a_k e^{-|k|y} + \left(1 - \frac{{\rm r}_{\rm H}}{{\rm r}^2}\right)b_k e^{-qy}\right]\, .
\end{equation}
The subject of our interest is the linear in ${\rm r_H}$ correction to the potential $\phi(x,y)$.

\subsection{General form of corrections}

In addition to the length scale ${\rm r}_H\ell$ associated with the Hall viscosity, the potential \eqref{eq_phi_k} depends on the length $\ell_s$ that characterizes the boundary conditions, and the length ${\rm r}\ell$ that depends on the dissipative regularization. In the QH regime discussed in the main text we have $\ell_s=\infty$ and ${\rm r}\ell\sim w$, but these parameters are not fixed in general, e.g.~in a dissipative regime. The potential will therefore have a complicated $x$ dependence, with crossovers when $x\sim \ell_s$ and $x\sim {\rm r}\ell$. However, unless there is an additional natural length scale in the system, it is natural for ${\rm r} \ell$ and $\ell_s$ to be either of order of the long distance cutoff $w$ (which is comparable to the size of the system) or the short distance cutoff $\ell$. The profile of the potential for $\ell \ll x \ll w$ is therefore considerably simpler. We will find below that it generically has the form
\begin{equation}\label{eq_phi_gen}
\phi(x,y) = \alpha_0\phi^{(0)}(x,y) + {\rm r_H}\left[\alpha_1 \phi^{(1)}(x,y) +  \alpha_2 \phi^{(2)}(x,y) + \ldots\right]\, ,
\end{equation}
where (writing $z=x+iy$)
\begin{equation}\label{eq_phi_n}
\phi^{(n)}(x,y) = -{\rm Re} \left[(-i)^{n+1} (\ell \d_z)^n\log z \right]\, .
\end{equation}
Note that when ${\rm r}\ell \to 0$, one can see from \eqref{eq_pde2} that $\phi$ is harmonic, and can thus be written as the real part of a holomorphic function. For ${\rm r}\ell$ finite, this holomorphic function is replaced by a Laurent series \eqref{eq_phi_gen} with a finite radii of convergence $\ell \lesssim |z| \lesssim w$.

The first few corrections are
\begin{equation}
\phi^{(0)} = \pi- {\rm Arg}(x+iy)\, , \qquad
\phi^{(1)} = \ell\frac{x}{x^2+y^2}\, , \qquad
\phi^{(2)} = \ell^2\frac{2xy}{(x^2+y^2)^2}\, , \qquad 
\phi^{(3)} = -\ell^3\frac{2(x^3- 3xy^2)}{(x^2+y^2)^3}\, , \quad \cdots \, .
\end{equation}
When ${\rm r_H}=0$, the only contribution to the potential is $\phi^{(0)}$, which is simply a step function at the boundary and leads to the usual quantized resistance of Hall systems. For most generic values of the parameters $({\rm r}\ell, \ell_s)$, all corrections $\phi^{(n)}$, $n>0$, will be present. The leading one is $\phi^{(1)}$ gives a contribution to the boundary potential $\phi(x,0)$. One exception is the case when ${\rm r \ell }\sim w$ and $\ell_s\sim \ell$ -- in this regime the leading contribution to the boundary comes from $\phi^{(3)}$. Finally, for certain special points of the parameter space $({\rm r}\ell, \ell_s)$ all corrections at the boundary vanish and one must instead measure the bulk correction $\phi^{(2)}$ to measure ${\rm r_H}$. These points are: (i) exact no-stress boundary condition $\ell_s=\infty$ (for both ${\rm r}\ell\sim \ell$ and ${\rm r}\ell\sim w$), and (ii) exact no-slip boundary conditions $\ell_s = 0$ when ${\rm r}\ell \sim w$. All cases are summarized in table \ref{tab_sum} below, and studied in the following sections.

\begin{table}[h]
\begin{center}
\begin{tabular}{|c||c|c|c|c|}
\hline
& $\ell_s = 0$  & $\ell_s \sim \ell$ & $\ell_s \sim w$ & $\ell_s =\infty$ \\
\hline\hline
$\vphantom{\displaystyle \int_1^1}{\rm r}\ell \sim \ell$ & Boundary $\phi^{(1)}\sim \dfrac{1}{x}$ & Boundary $\phi^{(1)}\sim \dfrac{1}{x}$ & Boundary $\phi^{(1)}\sim \dfrac{1}{x}$ & Near boundary $\phi^{(2)}$ \\
\hline
$\vphantom{\displaystyle \int_1^1}{\rm r}\ell \sim w$ & Bulk $\phi^{(2)}\sim \dfrac{1}{x^2}$ & Boundary $\phi^{(3)}\sim \dfrac{1}{x^3}$ & Boundary $\phi^{(1)}\sim \dfrac{1}{x}$ & Bulk $\phi^{(2)}\sim \dfrac{1}{x^2}$ \\
\hline
\end{tabular}
\end{center}
\caption{\label{tab_sum}Dominant correction due to ${\rm r_H}\neq 0$ to the potential near a contact where current is injected.}
\end{table}%

As discussed in the main text, in the QH regime only no-stress boundary conditions ($\ell_s=\infty$) are consistent with the fact that the Hall conductance in a pipe flow cannot receive corrections algebraic in the width of the pipe \cite{niu1987quantum}, we will therefore mostly focus on this case in the following. When $\ell_s = \infty$, all boundary corrections to the potential vanish and one must probe the bulk potential in order to measure the Hall viscosity or ${\rm r_H}$ (this explains the robustness of the quantization of the Hall conductivity in this more complex geometry).

\subsection{Bulk potential in QH regime}

When no-stress boundary conditions are imposed ($\ell_s=\infty$), only the potential in the bulk is sensitive to ${\rm r_H}$ and therefore to the Hall viscosity. We start by considering the case where ${\rm r}\ell\sim w$ (the other scenario ${\rm r}\ell \sim \ell$ is studied further below). The potential \eqref{eq_phi_k} is then given by
\begin{equation}\label{eq_phi_k_nostress}
\phi_k(y) = \frac{\rho_{\rm H}^{(0)}I}{ik} e^{-|k|y} \left[1+{\rm r_H}\ell^2 k^2\vphantom{\frac{}{}} + \ldots\right]\, ,
\end{equation}
where we are ignoring terms of order $k {\rm r}\ell \sim k w$ since we are interested in the potential $\phi(x,y)$ for $x,y\ll w$. Taking the Fourier transform, one finds
\begin{equation}\label{eq_phi_bulk}
\phi(x,y) = \rho_{\rm H}^{(0)} I \left[\frac{\pi -\theta(x,y)}{\pi} + \frac{2}{\pi} {\rm r}_{\rm H} \ell^2\frac{xy}{(x^2+y^2)^2} + \ldots \right]\, , 
\end{equation}
where $\theta(x,y) = {\rm Arg \,}(x+iy)$. Measuring the potential along the diagonal $x=y$, this is a correction to the background potential of the form $\phi_{(2)}\sim {\rm r_{\rm H}}\ell^2/x^2$. We will see below that with finite size contacts, the short-distance singularities in the potential as $x,y\to 0$ are smeared over the width of the contacts. This regulated form of the potential \eqref{eq_phi_bulk} was used to generate Fig.~\ref{fig_flow}.

We now turn to the regime of a short-distance dissipative regulator  ${\rm r} \ell\sim \ell$  (keeping no-stress boundary conditions $\ell_s = \infty$), this regime is labeled `Near boundary' in Table \ref{tab_sum}. Because of the no-stress boundary conditions, there is no boundary correction to the potential. Moreover, because ${\rm r}\ell$ is short-distance, the correction that is sensitive to ${\rm r_H}$ in \eqref{eq_phi_k} decays exponentially in the bulk. Specifically, it has the form
\begin{equation}
\delta\phi \sim \rho_{\rm H}^{(0)} I\, {\rm r_H}\ell^2 \frac{xy}{(x^2+y^2)^2}e^{-y/{(\rm r\ell)}}\, .
\end{equation}
This correction should be measured at a short but finite distance $\delta y$ away from the boundary, varying $x$.

\subsection{Finite size contacts}

Finite size contacts resolve the $x,y\to 0$ divergence in the potential \eqref{eq_phi_bulk}. The exact current distribution in the section of the contacts does not qualitatively affect this smoothing, so we will consider only rectangular and Gaussian for simplicity. The boundary condition \eqref{eq_bc_inject} is then changed to $\d_x\psi(x,0) = I f(x)$, with
\begin{equation}\label{eq_bc_inject_finite}
f(x) = \frac{\Theta(x+\frac{a}{2})\Theta(-x+\frac{a}{2})}{a}\quad \hbox{(Rectangle contacts)}\, , \qquad \hbox{or} \qquad
f(x) = \frac{e^{-x^2/a^2}}{a\sqrt{\pi}}\quad \hbox{(Gaussian contacts)}\, ,
\end{equation}
where $a$ is the width of the contact. As usual, a source can be added far away from the drain at $(x,y)=(w,0)$ for IR regulation, but we will not keep this regulator explicitly here. The solution to \eqref{eq_pde_k} is now
\begin{equation}
\psi_k(y) = \frac{I}{ik} \widetilde f(k) \left[a_k e^{-|k|y} + b_k e^{-qy}\right]\, ,
\end{equation}
where $a_k$ and $b_k$ were given above, and $\widetilde f$ is the Fourier transform of $f$ \eqref{eq_bc_inject_finite}. Using no-stress boundary conditions and assuming ${\rm r}\ell \gtrsim w$ (see Table \ref{tab_sum}), the potential in the bulk is then given by (writing $z=x+iy$)
\begin{equation}
\phi(z) =\rho_{\rm H}^{(0)}I \int_{-\infty}^\infty \frac{dk}{2\pi} \frac{e^{ikz}}{ik}\widetilde f(k) \left[1 + {\rm r_H}\ell^2 k^2 \vphantom{\frac{}{}}\right]\, .
\end{equation}
This integral can be performed exactly for either contact shapes in \eqref{eq_bc_inject_finite}. The potential along the $x=y$ line is plotted in Fig.~\ref{fig_finite_contact}, for rectangular contacts. For positive ${\rm r_H}$, the potential initially overshoots before settling down to its asymptotic value as $x=y\to \infty$. This signature is robust and survives even for contacts that are several factors larger than the characteristic length ${\rm r_H}\ell^2$.

It is possible to quantify how small the contacts must for ${\rm r_H}$ to be measurable. Expanding $\widetilde f(k) = 1 - \alpha (ak)^2 + O(ak)^4$ one finds that the potential for $a, {\rm r_H}\ell \ll x,y\ll w$ again has the form \eqref{eq_phi_bulk}, with the replacement
\begin{equation}
{\rm r_H \ell^2} \quad \to \quad 
{\rm r_H \ell^2} - \alpha a^2\, . 
\end{equation}
For rectangular contacts, $\alpha=1/24$, and for Gaussian contacts $\alpha=1/4$. Fig.~\ref{fig_finite_contact} shows how certain features in the potential survive even when the contact is several factors larger than $|{\rm r_H}|^{1/2}\ell$ -- for example, the potential `over-shoots' and reaches a maximum for finite $x$ if ${\rm r_H}>0$ and $a<\sqrt{24}|{\rm r_H}|^{1/2}\ell$ (for sharp rectangular contacts).

\end{widetext}

\end{document}